\documentclass[]{aa}

\usepackage{txfonts,epsfig,graphicx,natbib,url,twoopt}
\usepackage{textgreek}
\usepackage{import}
\usepackage{paralist}
\usepackage{booktabs,xspace}
\usepackage{amsmath}
\usepackage{tikz}
\usepackage{amsmath}
\usepackage{amssymb}
\usepackage{graphicx}
\usepackage{float}
\usepackage[export]{adjustbox}

\bibpunct{(}{)}{;}{a}{}{,}

\usepackage[hidelinks=true,breaklinks=true]{hyperref} %% to avoid \citeads line fills
\usepackage{natbib,twoopt}
\bibpunct{(}{)}{;}{a}{}{,} %% natbib format for A&A and ApJ
\makeatletter
 \newcommandtwoopt{\citeads}[3][][]{%
   \href{http://adsabs.harvard.edu/abs/#3}%
        {\def\hyper@linkstart##1##2{}%
         \let\hyper@linkend\@empty\citealp[#1][#2]{#3}}%   %% Rutten, 2000
	}
 \newcommandtwoopt{\citepads}[3][][]{%
   \href{http://adsabs.harvard.edu/abs/#3}%
        {\def\hyper@linkstart##1##2{}%
         \let\hyper@linkend\@empty\citep[#1][#2]{#3}}%     %% (Rutten 2000)
	}
 \newcommandtwoopt{\citetads}[3][][]{%
   \href{http://adsabs.harvard.edu/abs/#3}%
        {\def\hyper@linkstart##1##2{}%
         \let\hyper@linkend\@empty\citet[#1][#2]{#3}}%     %% Rutten (2000)
	}
 \newcommandtwoopt{\citeyearads}[3][][]{%
   \href{http://adsabs.harvard.edu/abs/#3}%
        {\def\hyper@linkstart##1##2{}%
         \let\hyper@linkend\@empty\citeyear[#1][#2]{#3}}%  %% 2000
	}
\makeatother

\renewcommand{\vec}{\boldsymbol} %bold italic for vectors

\begin{document}
	\title{Quasars can be used to verify the parallax zero-point of the Tycho--Gaia Astrometric Solution}
	\titlerunning {Quasars can be used to verify the parallax zero-point of TGAS}
	\authorrunning {D.~Michalik and L.~Lindegren}
	\author{
		Daniel Michalik
			\and
		Lennart Lindegren
	}

	\institute{
		Lund Observatory, Department of Astronomy and Theoretical Physics, Lund University, Box 43, 22100 Lund, Sweden\\
		\email{[daniel.michalik; lennart]@astro.lu.se}
	}

	\date{Received 24 September 2015 / Accepted 25 October 2015}

	\abstract
{The Gaia project will determine positions, proper motions, and
parallaxes for more than one billion stars in our Galaxy. It is known that
Gaia's two telescopes are affected by a small but significant variation of the
basic angle between them. Unless this variation is taken into account 
during data processing, e.g. using on-board metrology, it causes systematic
errors in the astrometric parameters, in particular a shift in the parallax
zero-point. Previously, we suggested an early reduction 
of Gaia data for the subset of Tycho-2 stars (Tycho-Gaia Astrometric Solution; TGAS).}
{
We investigate whether quasars can be used to independently verify the
parallax zero-point in early data reductions. This is not
trivially possible as the observation interval is too short to disentangle parallax and
proper motion for the quasar subset.
}
{We repeat TGAS simulations but additionally include simulated Gaia
observations of quasars from ground-based surveys. All observations are
simulated with basic angle variations. To obtain a full astrometric solution
for the quasars in TGAS we explore the use of prior information for their
proper motions. 
}
{ 
It is possible to determine the parallax zero-point for the quasars with a few
\textmu as uncertainty, and it agrees to a similar precision with the zero-point
for the Tycho-2 stars. The proposed strategy is
robust even for quasars exhibiting significant spurious proper motion due to a
variable source structure, or when the quasar subset is contaminated with stars
misidentified as quasars.
}% conclusions heading (optional), leave it empty if necessary
{
Using prior information about quasar proper motions we could provide an independent verification of
the parallax zero-point in early solutions based on less than one year of Gaia
data. 
}

\maketitle

% ==============================================================================
% ==============================================================================
\section{Introduction\label{sec:introduction}} 
The European space mission Gaia determines astrometry, photometry,
and spectroscopy for more than one billion sources\footnote{The word source
refers to any point-like object observed by
Gaia; this includes stars, quasars, supernovae, etc.}
\citep{2001A&A...369..339P,2012Ap&SS.341...31D}. Important features of
Gaia's astrometric measurements are 
\begin{itemize}
\item the uniform scanning that ensures a relatively homogeneous all-sky performance; 
\item the high accuracy of the final astrometric data, at a level of tens of \textmu as for $G=15$;
\item the relatively faint $G\simeq 20$ magnitude
limit, which makes it possible to observe a large number of quasars, necessary
for the determination of the reference frame and as an independent check of the
parallax zero-point;
\item and the capability to measure absolute parallaxes 
by combining simultaneous measurements of different objects separated by a large angle on the sky.
\end{itemize}
For the last point, Gaia's design includes two viewing directions separated by a large basic angle, which
needs to be either perfectly stable or independently monitored. Gaia's basic angle was designed to be very stable, while
at the same time being measured on board with high accuracy through an
interferometric device called the basic angle monitor (BAM; \citealt{2014SPIE.9143E..0XM}).

Verification of the stability of the basic angle and of the quality of the
on-board metrology can be done only partially through the analysis of the
post-fit residuals of the astrometric solution;  a full verification
requires the use of external data. Quasars provide a clean and self-consistent
approach, as they are so far away that their true parallaxes can safely be
assumed to be zero. It is thus possible to determine the zero-point of the
parallaxes measured by Gaia simply by taking the median of the resulting
parallax distribution in a quasar subset and comparing it to the expected zero
value. The width of this distribution gives an indication of the uncertainty
of the obtained median value.

For a full five-parameter solution of the astrometric parameters (position,
parallax, and proper motion), at least five distinct observations of each
source are necessary, unless prior knowledge can be used to complement the
observational data (\citealt{2015arXiv150702963M}). A full five-parameter data
reduction with less than one year of Gaia data is possible, for example, for the
Tycho-2 (Tycho--Gaia Astrometric Solution; TGAS;
\citealt{2015A&A...574A.115M}) and the Hipparcos stars (Hundred Thousand
Proper Motions project; HTPM; \citealt{LL:FM-040,2014A&A...571A..85M}).
The Tycho-2 and Hipparcos
catalogues contain extremely few extragalactic objects, which are not
sufficient for an independent verification of the basic angle.  Adding quasars
to such early solutions requires prior information to overcome the ambiguity of
parallax and proper motion. In this paper we explore which prior information
can be used, and demonstrate the feasibility of adding quasars to the TGAS
project for verification of the parallax zero-point in the light of basic
angle variations.

% ==============================================================================
% ==============================================================================
\section{Basic angle variations and metrology}
\begin{figure}[t]
\centering
\includegraphics[width=.35\textwidth,clip,trim=0 120 0 0]{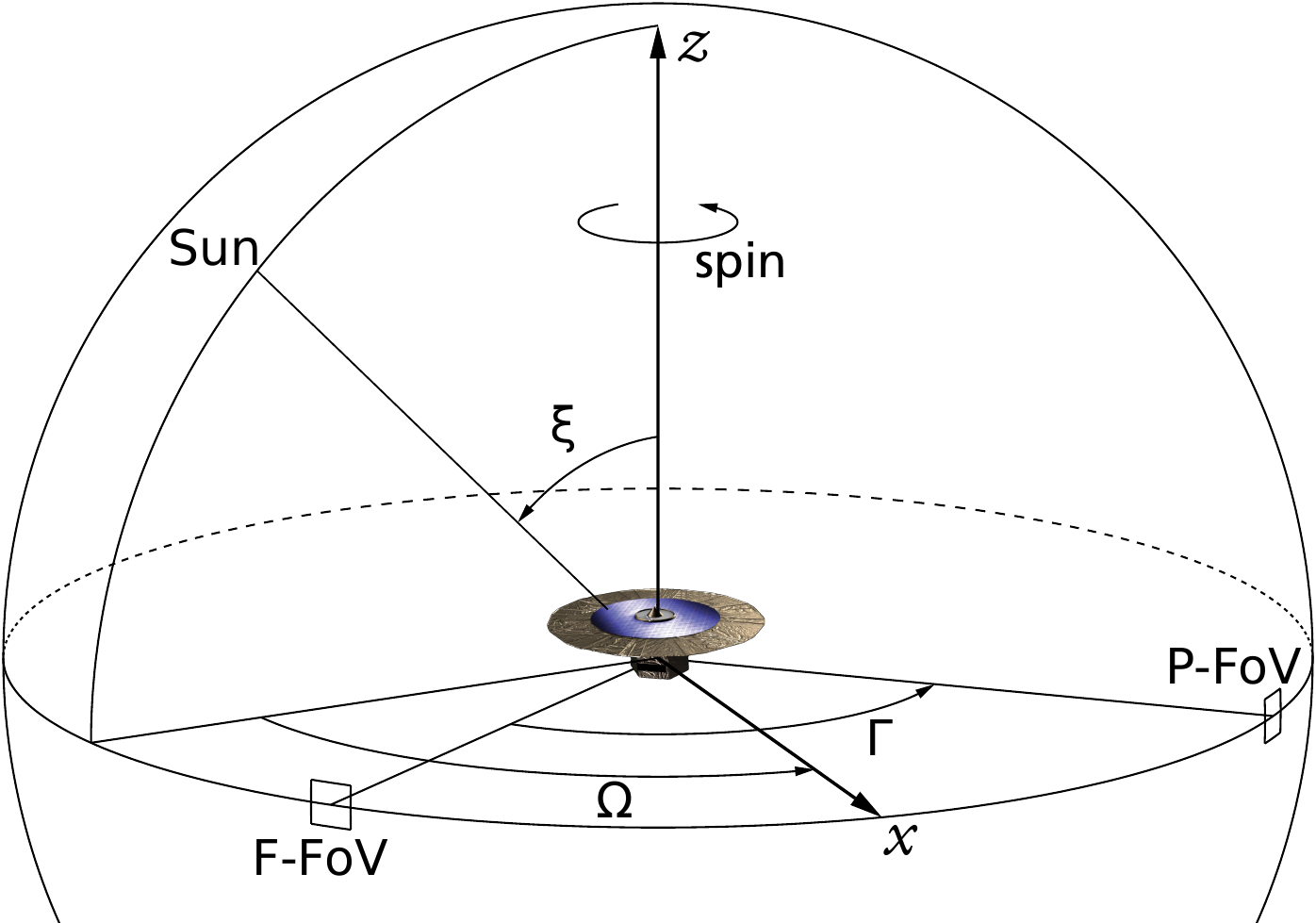}
\caption{Solar-aspect angle $\xi$ and spin phase $\Omega$ define the orientation of the spacecraft relative to the Sun. $\vec{x}$ and $\vec{z}$ are axes fixed in the spacecraft reference system. The basic angle $\Gamma$ separates the preceding and following fields of view (FoV), with  
$\vec{x}$ half-way between them.\label{fig:angles}}
\end{figure}
The two viewing directions of Gaia are separated by a basic angle $\Gamma =
106.5 \deg$. Basic angle variations are harmful to the resulting
astrometry unless they are modelled as part of the data processing 
or corrected by means of data from the on-board metrology device, BAM. The BAM deploys a laser beam to create an interferometric pattern in
each field of view (FoV). Variations in the basic angle cause a change in the
relative phases of the fringes, which are measured by a dedicated BAM CCD
adjacent to the main astrometric field of Gaia. It is
desirable to verify that these measurements correctly characterize the
variations for the entire focal plane. This can be done by comparing the BAM data with
the variations determined from the astrometric observations themselves.

Gaia's scanning requires a constant tilt $\xi = 45\deg$ of the spacecraft spin axis with
respect to the Sun \citep{2012Ap&SS.341...31D}. The phase of the spacecraft
relative to the Sun is therefore completely described by the angle $\Omega(t)$
giving the pointing of the satellite within its six-hour spin period (Fig.~\ref{fig:angles}).
\citet{2014SPIE.9143E..0XM} reported an early analysis of BAM measurements finding
stable periodic variations depending on $\Omega$ with an amplitude of about 1000~\textmu as.
This is much larger than expected from the design of the spacecraft;
however, the effects on the astrometric results can be largely eliminated if the
basic angle variations are determined with sufficient accuracy. The basic angle
variations can be described by a Fourier expansion in terms of $\cos k\Omega$
and $\sin k\Omega$ where $k=1,2,\ldots$ is the order of the harmonics. 

Dedicated simulations have shown that 
all but the $\cos \Omega$ term can be solved with high accuracy using Gaia data
alone, even with less than one year of observations. They are thus neglected throughout the rest of this article.
However, the first cosine harmonic is virtually 
indistinguishable from a constant shift of the parallax zero-point
\citep[Sect.~6.1]{1992A&A...258...18L} and therefore impossible\footnote{It has
been suggested that the finite size of the FoV and other design details of Gaia
may allow  even the $\cos \Omega$ term to be determined purely from the observational data
(S.~Klioner, private communication).} to determine from Gaia
data alone. \citet{LL:GAIA-LL-057} relates the parallax zero-point $\Delta \varpi$ to
the amplitude $a_1$ of the $\cos \Omega$ term and the spacecraft distance $R$ (in au) from the Sun as
\begin{equation}
\Delta \varpi = \frac{a_1}{2R\sin\xi\sin(\Gamma/2)}.\label{eq:shift}
\end{equation}
In this paper we limit the further analysis to the first cosine term assuming a
fixed amplitude $a_1 = 1000$~\textmu as. For the observation interval used in the
following simulations $R$ evolves such that the expected average is $\Delta \varpi \simeq 871.9$~\textmu as.

% ==============================================================================
\section{Quasar parallaxes in early solutions}
We first repeat simulations of the TGAS scenario as described in \citet[][
Sect.~3]{2015A&A...574A.115M}, but perturb the observations by a 
periodic basic angle variation proportional to
$\cos\Omega$ with an amplitude of 1000~\textmu as. Otherwise we follow the same 
assumptions as before, i.e.\ we simulate half a year of Gaia
observations of the Hipparcos and Tycho-2 stars, and process them using the
Hipparcos positions and proper motions and the Tycho-2 positions at 1991.25 as
priors. As expected, the resulting parallax solution is
strongly biased with a median parallax error (estimated minus true value)
consistent with Eq.~(\ref{eq:shift}). This zero-point shift cannot be
easily determined from the stellar observations themselves, and the recovery of
absolute parallaxes in such a solution must instead rely on the
correctness of BAM metrology, which can be verified by using external information.

Thus, it is desirable to include an additional subset of quasars. The
true quasar parallaxes are known to be virtually zero. Therefore, the median of the quasar
subset can be used to estimate the parallax zero-point of the astrometric
solution. For our simulations the quasar subset is taken from the Gaia Initial Quasar Catalogue (GIQC;
\citealt{2014jsrs.conf...84A}), which is a list of quasars produced in preparation for the
Gaia mission. It is based mainly on the Large Quasar Astrometric Catalogue (LQAC;
\citealt{2012A&A...537A..99S}), which itself is based on the Sloan Digital Sky Survey (SDSS; \citealt{2009ApJS..182..543A})
and other ground-based surveys. It contains positions and approximate magnitudes for
over one million objects. 
 The source distribution is strongly inhomogeneous and
shows the survey footprint. 
Within the magnitude limits of Gaia we use the $\sim$190\,000
entries flagged as ``defining'', i.e.\ objects that are quasars with a high level of certainty based on their observational
history and spectroscopic properties. 

To account for the possibility that early
Gaia solutions might not include observations for all of them, 150\,000
($\sim$80\%) of the quasars listed in GIQC are randomly selected and the rest are
discarded. From GIQC we use the position and magnitude to define the simulated `true'
quasar sources. The true values of parallax and proper motion are initially set to zero.

To allow us to obtain a sensible five-parameter solution with a
stretch of data as short as in a half-year TGAS solution, we need some prior
information for the quasars. One could consider using the precisely known radio
positions of VLBI quasars. Approximately 2500 ICRF sources with optical
counterparts are expected to be bright enough to be detected by Gaia. However,
this number of sources is too small to provide a statistically meaningful result. For the much larger
number of GIQC quasars no reliable position information exists at a level that
makes it usable as prior information. Even though we know that quasar
parallaxes are supposed to be zero, we do not want to use this as prior
information in the solution either, since we want to determine the parallax
values freely from the Gaia data in order to verify the parallax zero-point.
Instead, we suggest making use of the fact that quasars have negligible proper motions%
\footnote{Spurious proper motions caused by intrinsic variations in the quasars are discussed in Sect.~\ref{sec:spurious}.
Additionally, the expected proper motion of the Galactocentric acceleration must be taken into account in the real data. 
This effect is a few \textmu as~yr$^{-1}$ \citep{1995ESASP.379...99B,2003A&A...404..743K} and does not affect 
the principle shown in this paper.}
due to their cosmological distances.
Incorporating this information as a prior in the early
Gaia astrometric solutions will lift the parallax--proper motion degeneracy and is sufficient to obtain a good
astrometric solution for the quasar subset.

We demonstrate the feasibility of the method through three different simulations. 
First we use a clean quasar sample with zero true proper motions and
parallaxes. Then we relax these assumptions and introduce quasar structure
variations, as well as contamination of the dataset with stellar sources.
Table~\ref{tab:results} shows the results of the three experiments (see below for  further
explanations).

\subsection{Clean quasar sample}
In the first experiment the simulated true parallaxes and proper motions
in the quasar subset are strictly zero. To allow a full five-parameter
astrometric solution we apply a prior of $0 \pm 10$~\textmu as~yr$^{-1}$ to each
proper motion component. The prior uncertainty of 10~\textmu as~yr$^{-1}$ is
somewhat arbitrary, but provides enough weight to constrain the proper motions
to negligible values without causing numerical difficulties. We incorporate the
prior using Bayes' rule as described in \citet{2015arXiv150702963M}.

We evaluate the resulting parallaxes separately for the stellar subset (Hipparcos and 
Tycho-2 stars) and the quasars. Table~\ref{tab:results}, experiment~1, presents the median value
of the parallax errors (estimated minus true), the uncertainty of the median calculated using the bootstrap
method, and the RSE%
\footnote{ The ``robust scatter estimate'' (RSE) is defined as 0.390152 times
the difference between the 90th and 10th percentiles of the distribution of the
variable. For a Gaussian distribution it equals the standard deviation. Within
the Gaia core processing community the RSE is used as a standardized, robust
measure of dispersion \citepads{2012A&A...538A..78L}.\label{footnote:RSE}}
dispersion of the parallax errors for each of
the subsets. The different columns give statistics for selections based on 
the individual formal standard uncertainties of the parallaxes. The median
obtained for the quasar subset agrees
with the corresponding stellar value to within a few \textmu as, independent of
the selection of sources. 

\begin{table}[t]
\small
\centering
\caption{Simulation results of three different experiments comparing the
parallax median between the stellar subset and the quasars.
\label{tab:results}}
\begin{tabular}{lrrrr}
\toprule[\arrayrulewidth] \toprule[\arrayrulewidth]
% inserts double horizontal lines with same thickness as hrule. toprule is better than hrule since it gets the vertical spacing right.
& \multicolumn{4}{c}{Parallax selection}\\
\cmidrule[0.2pt](lr){2-5}
 & \multicolumn{2}{c}{90\% best} & \multicolumn{2}{c}{all}\\
\cmidrule[0.2pt](lr){2-3}
\cmidrule[0.2pt](lr){4-5}
Subset & Median [\textmu as]& RSE [\textmu as]&  Median [\textmu as]& RSE [\textmu as]\\
\midrule[\arrayrulewidth]  % inserts single horizontal line of same thickness as hrule. midrule is better than hrule since it gets the vertical spacing right.
\multicolumn{5}{c}{Experiment 1: clean quasar sample}\\
Stars		&  872.1  $\pm$ 0.2   &   441.9 & 872.1  $\pm$ 0.2   &   613.5 \\
Quasars	  	&  876.4  $\pm$ 2.0   &   1336.6 & 876.7  $\pm$ 2.5   &   2324.7 \\
\midrule[\arrayrulewidth]  % inserts single horizontal line of same thickness as hrule. midrule is better than hrule since it gets the vertical spacing right.
\multicolumn{5}{c}{Experiment 2: with spurious proper motions}\\
Stars		&  872.0  $\pm$ 0.2   &   442.0 & 872.0  $\pm$ 0.2   &   613.4 \\
Quasars	  	&  876.7  $\pm$ 2.9   &   1644.7 & 877.7  $\pm$ 3.4   &   2676.3 \\
\midrule[\arrayrulewidth]  % inserts single horizontal line of same thickness as hrule. midrule is better than hrule since it gets the vertical spacing right.
\multicolumn{5}{c}{Experiment 3: with 5\% contamination}\\
Stars		&  872.1  $\pm$ 0.2   &   441.9 & 872.0  $\pm$ 0.2   &   613.5 \\
Quasars 	&  871.7  $\pm$ 2.2   &   1429.2 & 872.0  $\pm$ 2.4   &   2452.9  \\
\bottomrule[\arrayrulewidth]  % inserts single horizontal line with same thickness as hrule uses. bottomrule is better than hrule since it gets the vertical spacing right.
\end{tabular}
\tablefoot{``Stars'' refers to the combined subset of Hipparcos and Tycho-2
sources. In each subset, statistics are given for the selection of 90\% of the sources with the
smallest individual formal uncertainties and for all sources together.
The values given are the median (and its uncertainty from the bootstrap method) and the RSE dispersion of the parallax errors (estimated minus true).}
\end{table} 
\subsection{Spurious proper motion from variable source structure\label{sec:spurious}}
Variation in the source structure of quasars can lead to shifts of their
photocentres up to the milliarcsecond level
\citep[e.g.][]{2012A&A...538A.107P,2009A&A...505L...1P,2011A&A...526A..25T}.
Linear trends of these shifts might lead to spurious proper motions measured
for quasars and stable over years to decades.
\citet{2011A&A...529A..91T} fitted long-term proper motions for 555 quasars from VLBI observations.
The total proper motion $\mu = \sqrt{\mu_{\alpha *}^2 + \mu_{\delta}^2}$ 
in $\mu$as~yr$^{-1}$ in their catalogue can be described by a log-normal distribution with mean
1.9~dex and standard deviation 0.61~dex. 
It is impossible to say whether these measurements give
an optimistic or conservative characterization of spurious quasar proper
motions on the much shorter time baselines of our simulations.
Additionally, the morphology of the host galaxy might lead to a statistical
increase in the centroiding error, and photometric variability of the nucleus
together with the stable photocentre of the host galaxy might lead to an 
effect similar to ``variability-induced movers'' in binaries \citep{1996A&A...314..679W}.
Physically all of these effects are expected to be random and therefore should
only increase the dispersion of the results but not the median values themselves.

We use the statistical properties of the results by \citet{2011A&A...529A..91T} as the basis for
simulations, but apply a factor of 10 to provide a conservative assumption on the total spurious motion.
The individual components of the proper motion are computed as
\begin{equation}
\mu_{\alpha *} = \mu \sin\theta,\quad
\mu_{\delta} = \mu \cos\theta,
\end{equation}
where $\theta$ is a random position angle and $\log_{10}\mu$ is taken from a
normal distribution with mean value 2.9~dex and standard deviation 0.61~dex.
The median value of the resulting $\mu$ is about 800~\textmu as~yr$^{-1}$.  While
this spurious proper motion increases the RSE of the solution for the quasar
subset, the agreement of median parallax between the quasars and the stellar
subset remains at the previous level (see Table~\ref{tab:results},
experiment~2).  This shows that significant spurious proper motions due to
photocentre variability do not harm the proposed strategy. 

\subsection{Contamination through misidentification}
One potential problem with the use of quasars for the zero-point verification
will be the identification of quasars in the Gaia observations. It is possible
that a small fraction will be misclassified. Stars
mistaken for quasars may have a noticeable parallax and proper motion which
could contaminate the results obtained for the presumed quasar subset. To
characterize the deterioration caused by misclassification, we replace 5\% of
the quasars by stellar sources. We assume that misclassification will be
most prevalent for faint sources where no good spectra exist, and obtain true
positions, parallaxes, and proper motions for contaminating stars from the Gaia Universe Model Snapshot (GUMS; \citealt{2012A&A...543A.100R}). We
use the 7500 brightest stars fainter than magnitude 19. The results for experiment 3 in
Table~\ref{tab:results} present the combined evaluation of the quasar subset
including the contaminating stars. Even with the contamination the median
parallaxes of the quasar subset still agree to within a few \textmu as
with the values found for the other subsets. 

\section{Conclusions} 
We present a strategy to verify the parallax zero-point in a TGAS solution in
the presence of basic angle variations. It uses quasars, which can only be included in the solution 
if prior information is applied. In the absence of accurate prior position information --
available only for a small number of VLBI quasars -- we propose to constrain 
their proper motions. Simulations show that this allows us to recover the parallax
zero-point in a solution with half a year 
of Gaia data to within a few \textmu as. This is true even if the
quasars exhibit considerable variability in their photocentres, provided the resulting spurious proper motions are
random from source to source. Furthermore, the scheme is robust to the quasar subset being
contaminated by a significant fraction of stellar sources misclassified as quasars. In all
cases the zero-point determined from the quasars agrees well with the
theoretically expected parallax shift from the basic angle perturbations applied in the
simulations.

Practical difficulties using quasars may arise from the colour calibration of
the point spread function, which is based on stellar sources. Quasars, however,
have very different spectra, which may require a separate calibration
(U.~Bastian, private communication). Whether this can be overcome in practice
remains to be seen.

%______________________________________________________________
% Acknowledgements and References should come before Appendices

\begin{acknowledgements}
Uli Bastian, Anthony Brown, Jos de Bruijne, David
Hobbs, Sergei Klioner, Timo Prusti, and an anonymous referee gave helpful
comments on the draft manuscript, for which we are grateful. Thanks to
Alexandre Andrei for
providing the GIQC data and information about its content. This work was
supported by the Gaia Data Processing and Analysis Consortium (DPAC) and uses
the AGIS/AGISLab and GaiaTools software packages; special thanks to their
maintainers and developers. We gratefully acknowledge financial support from
the Swedish National Space Board, the Royal Physiographic Society in Lund, and
from the European Space Agency (contract no.~4000108677/13/NL/CB). Quasar
proper motion data were obtained through VizieR.
\end{acknowledgements}

\bibliographystyle{aa} % style aa.bst
\bibliography{quasarsInTGAS} % your references Yourfile.bib

\appendix

\end{document}